\documentclass[twocolumn,showpacs,preprintnumbers,amsmath,amssymb]{revtex4}

\usepackage{graphicx}
\usepackage{dcolumn}
\usepackage{bm}
\usepackage{natbib}
\begin{document}

\preprint{Phys.\ Rev.\ Lett. {\bf 100}, 017208 (2008)}

\title{Topology of the Spin-polarized Charge Density in bcc and fcc Iron}
\author{Travis E. Jones$^\dagger$}
 \email{trjones@mines.edu}
\author{Mark E. Eberhart$^\dagger$}
\author{Dennis P. Clougherty$^{\dagger\ddagger}$}

\affiliation{$^\dagger$ Molecular Theory Group,
Department of Chemistry and Geochemistry,
                         Colorado School of Mines,
                         Golden, Colorado 80401 }

\affiliation{$^\ddagger$ Department of Physics,
                   University of Vermont,
                   Burlington, Vermont 05405-0125 }

\date{July 9, 2007}

\begin{abstract}
We investigate the topology of the spin-polarized charge density in bcc and fcc iron. While the total spin-density is found to possess the topology of the non-magnetic prototypical structures, in some cases the spin-polarized densities are characterized by  unique topologies; for example, the spin-polarized charge densities of bcc and high-spin fcc iron are atypical of any known for non-magnetic materials.  In these cases, the two spin-densities are correlated: the spin-minority electrons have directional bond  paths with deep minima in the  minority density, while the spin-majority electrons  fill these holes, reducing bond directionality. The presence of two distinct spin topologies suggests that a well-known magnetic phase transition in iron  can be fruitfully reexamined in light of these topological changes. We show that the two phase changes seen in fcc iron (paramagnetic to low-spin and low-spin to high-spin) are different. The former follows the Landau symmetry-breaking paradigm and proceeds without a topological transformation, while the latter also involves  a topological catastrophe.
\end{abstract}

\maketitle


Bader's topological theory of molecular structure, Atoms in Molecules (AIM), has been successfully applied to a variety of crystalline systems \cite{AIM, Zou:1994, Luana:1997,Eberhart:1998,Bond_bundle,Tsirelson:1995,Eberhart:1993, Eberhart:2001,Eberhart:1996c,Eberhart:1996b, Kioussis:2002}. It has been employed to investigate the nature of bonding in materials ranging from high temperature alloys to biological systems. These have yielded surprising results, such as second neighbor bond paths in B2 ionic crystals \cite{Luana:1997} and transition metal aluminides \cite{Eberhart:1998}, with the magnitude of the latter correlating to failure properties \cite{Bond_bundle}. Other studies have used bond path properties to offer first principles explanations of stress-induced failure in brittle and ductile alloys \cite{Eberhart:1993}, as well as shear elastic constants in a variety of pure metals and alloys \cite{Eberhart:2001,Eberhart:1996c,Eberhart:1996b}. Building on these ideas and using the rigorous definitions of bond paths afforded by the theory, the anomalous behavior of iridium under shear was also explained \cite{Kioussis:2002}. 

One of the attractive features of AIM is its reliance on the charge density, a quantum mechanical observable, that is most often calculated but can, in principle, be measured via X-ray diffraction techniques \cite{Arnold:2000,Farrugia:2005,Lyssenko:2004,Koritsanszky:2001}.  In a similar fashion, the spin-polarized charge density is an observable that can be calculated or measured using spin-polarized neutron diffraction.  Despite the information and insights that have come from topological investigations of the total charge density, the same analysis has yet to be performed on spin-polarized densities.   Here, we report the results from the first such studies, exploring the spin-minority and spin-majority topologies of body-centered-cubic (bcc) and face-centered-cubic (fcc) iron. 

This first application of AIM to spin-density sheds light on the origins of the magnetic phase transitions of fcc iron. It is argued that this system undergoes two distinct phase transitions during volume expansion: a second-order phase change, from a paramagnetic to a low-spin state, occurs at smaller volumes. It is coincident with a change in the charge density at the critical points without a topological transformation. At larger volumes, the low-spin state changes to high-spin through a discontinuity in the moment, where the topology also transforms.

The framework of AIM can be applied in the same manner to both the total and spin-polarized densities. It is known from the Hohenberg-Kohn theorem that all ground-state molecular properties are a consequence of its charge density $\rho(\vec{r})$ \cite{HK}, a scalar field in three spatial dimensions. Bader noted $\rho(\vec{r})$ must also contain the essence of a molecule's structure, which can be described topologically.

The topology of a general scalar field can be characterized in terms of its critical points (CPs), the zeros of the gradient of this field. There are four kinds of CP in a three-dimensional space: a local minimum, a local maximum and two kinds of saddle points. These CPs are conventionally denoted by an index, which is the number of positive curvatures minus the number of negative curvatures. For example, a minimum CP has positive curvature in three orthogonal directions; therefore it is called a (3, 3) CP. The first number is simply the number of dimensions of the space, and the second number is the net number of positive curvatures. A maximum is denoted by (3, -3), since all three curvatures are negative. A saddle point with two of the three curvatures negative is denoted (3, -1), while the other saddle point is a (3, 1) CP. 

Through extensive studies of molecules and crystals, Bader and Zou and Bader showed that it was possible to correlate topological properties of the charge density with elements of molecular structure and bonding \cite{AIM,Zou:1994}. In particular, a bond path was shown to correlate with the ridge of maximum charge density connecting two nuclei, such that the density along this path is a maximum with respect to any neighboring path. The existence of such a ridge is guaranteed by the presence of a (3, -1) CP between bound nuclei. As such, this CP is sometimes referred to as a bond CP. 

Other types of CPs have been correlated with other features of molecular structure. A (3, 1) CP is topologically required at the center of ring structures, e.g. benzene. Accordingly, it is designated a ring CP. Cage structures are characterized by a single (3, 3) CP and again are given the descriptive name of cage CPs.  The locations of the atomic nuclei always coincide with a maximum, a (3, -3) CP. Hence it is conventionally called an atom CP.

In this study, iron's spin-polarized charge densities were calculated using the Vienna {\it ab-initio} simulation package (VASP) version 4.6 \cite{VASP1,VASP2}, which is based on the projector augmented wave method (PAW) \cite{Blochl,PAW}. The Perdew-Wang (PW91) generalized gradient corrections \cite{PW91} and the Vosko-Wilk-Nusair interpolation \cite{VWN, dpc91, Clougherty:1991} of the correlation part of the exchange-correlation functional were included.

Calculations performed on non-magnetic, monotonic, bcc transition metals show that the ground state total charge densities share the same topology as the prototype structure, W. Each atom has eight bonds to the nearest neighbor atoms, with no known case of second neighbor bond paths in elemental bcc materials. This also holds true for the total charge density of ferromagnetic bcc ($\alpha$) Fe; however it does not hold for the spin-density topology.

\begin{figure*}
\includegraphics{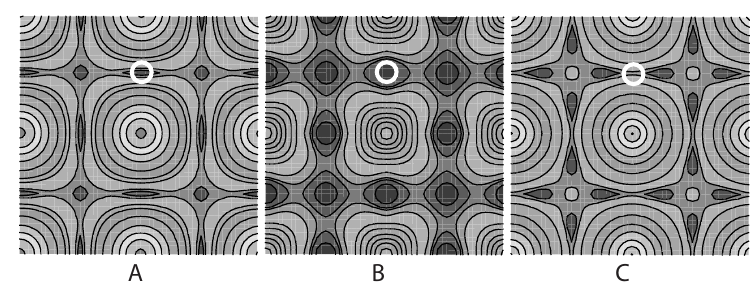} 
\caption{\label{fig:fig1}Total (A), spin-minority (B), and spin-majority (C) charge density contour plots in bcc Fe. The cut-plane passes through second neighbor atoms. Darker regions correspond to low charge density. Cage points, circled in white, are present in the total (A) and spin minority (B) charge densities. Second neighbor bond points (one circled in white) replace these in the spin-majority (C) charge density.}
\end{figure*}

Two phases of bcc Fe with lattice constants (moments) of 2.87 \AA  (2.22 $\mu_B$) \cite{Villars:1985,Danan:1968} and 2.83 \AA (2.21 $\mu_B$), respectively, were investigated, yielding identical topologies. The total charge density topology is prototypical bcc, as can be seen in  Fig.~\ref{fig:fig1}A, where a contour plot through second neighbor atoms in a (100) plane is shown.  A cage point, circled in Fig.~\ref{fig:fig1}A, is present.  Similarly, the spin-minority density reflects the topology of the total, Fig.~\ref{fig:fig1}B.  When the spin-majority density is examined, however, a different picture emerges.  Now, instead of a cage point between second neighbor atoms,  a bond point is found,  Fig.~\ref{fig:fig1}C, giving rise to 14 bond paths in the spin-majority topology.

Our results are consistent with the band theory of ferromagnetism, wherein, the{ \it{d}}-band splits into bonding, spin-minority, and less bonding, spin-majority, bands.  As bcc Fe has $O_{h}$ symmetry about the atoms, the d-orbitals reduce as $T_{2g}$ and $E_g$ at the band $\Gamma$-point.  The spin-minority electrons predominately populate the triply degenerate state, to produce 8 $\sigma$-bonds directed towards cube corners. The electrons in the spin-majority band are more equitably distributed between the $T_{2g}$ and $E_g$ bands. The inclusion of electrons of $E_g$ symmetry increases the charge density at the cube faces, e.g., second neighbor bond paths are formed.  These are, however, weakly directional.  

Directionality  has been quantified by two components, using the quadratic surface constructed from the Hessian of the charge density (or spin-polarized charge density) at the bond CP \cite{Eberhart:1996c,Eberhart:1996b,Kioussis:2002, Eberhart:1998}.  An elliptic cone surrounding the bond path is a quadratic surface consistent with the signs of the principal curvatures at the bond CP.  The extreme angles of the cone with respect to the plane normal to its axis are given by 
\begin{equation}
\tan\alpha=\sqrt{\rho_{\perp\perp}\over\displaystyle \rho_{\parallel\parallel}}
~{\rm{and}}~
\tan\beta=\sqrt{\rho_{\perp'\perp'}\over\displaystyle \rho_{\parallel\parallel}}
\label{direction}
\end{equation}
where $\rho_{\parallel\parallel}$ is the principal curvature parallel to the bond path and $\rho_{\perp\perp}$ and $\rho_{\perp'\perp'}$ are the two principal curvatures perpendicular to the bond path, which are degenerate in the bcc structure. This notation is employed to reflect the fact that the terms are all second derivatives of $\rho(\vec{r})$ in the direction indicated by the subscripts.   The definition in Eq.~\ref{direction} is trivially generalized to directionality for the spin-polarized charge density by affixing a spin index.

This definition  was motivated by the desire to associate a ``distance'' to bond breaking, i.e. the vanishing of a bond CP and the transformation of the elliptic cone to intersecting planes. This will occur when one (or both) of the principal curvatures perpendicular to the bond path vanishes.  By definition, then, bonds with larger values of directionality are further from a topological instability.  The degenerate components of the spin-minority bond directionality are found to be 0.565, while those of the spin-majority are 0.472 and 0.227 for the first and second neighbor bond paths, respectively.  In comparison, the bond directionality is 0.597 in W and  0.502 in bcc Fe.

\begin{figure}
\includegraphics{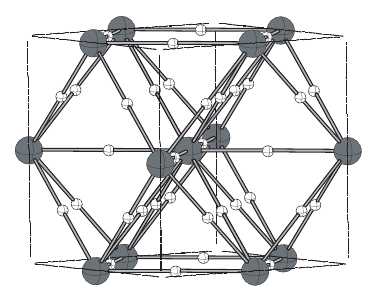} 
\caption{\label{fig:fig2} Topology of fcc prototype. Only bond points (white) and maxima (gray) are shown for clarity. Paramagnetic, LS  and the total and spin-majority density in HS fcc Fe display this topology. (Though not shown, cage points are present in the tetrahedral holes.)}
\end{figure}

When non-magnetic fcc metals were examined, only first neighbor bond were found in the total charge density, Fig.~\ref{fig:fig2}. Each atom has a total of 12 bond paths, and both the octahedral and tetrahedral holes contain a cage point.  However, as with magnetic bcc Fe, the spin densities give rise to distinct topologies.  In the case of fcc ($\gamma$) Fe, three magnetic phases are observed, high-spin, low-spin, and paramagnetic (HS, LS, and PM respectively) phases, each characterized by a different range of lattice constants \cite{Krasko:1987,Kaufman:1963,Moruzzi:1989,Bagayoko:1983,Kong:2004,Kubler:1981}.  From the VASP calculations we find the HS phase corresponds to $a\ge 3.564$ \AA, while the LS, low volume phase, occurs for $a\le3.563$  \AA, with the PM phases equilibrating at $a=3.45$ \AA. 

The LS/HS phase change proceeds through a magnetovolume instability \cite{Moruzzi:1989,Bagayoko:1983}. As the lattice parameter is increased from its LS phase ($a=3.465$ \AA) value, the moment increases smoothly from 0.75 $\mu_B$ to 1.3 $\mu_B$ at $a=3.563$ \AA. Then over a 0.001 \AA~increase in lattice constant a near doubling of the moment to 2.5 $\mu_B$ at $a=3.564$ \AA~ results.  Our results agree well with other computational \cite{Kong:2004,Kubler:1981} and experimental studies \cite{Mitani:1993,Macedo:1998}. Less clear is the PM/LS phase change. The situation yields ambiguous results if $\mu$ is used as an order parameter, with the transition appearing as first \cite{Krasko:1987,Moruzzi:1989} or second \cite{Kong:2004} order. Small changes in total energy associated with this phase change, 0.001 eV/atom \cite{Kong:2004}, make the correct assignment of moment problematic. 

The topology of the PM phase is identical to the fcc prototype, where $\alpha$ and $\beta$ are not equivalent.  In the fcc structure, the directions of principal curvature perpendicular to the bond path at the bond CP point toward the octahedral and tetrahedral holes are denoted as   $\rho_{oo}$ and $\rho_{tt}$ respectively.  Now, the components of directionality become
\begin{equation}
\tan\alpha=\sqrt{\rho_{oo}\over\displaystyle \rho_{\parallel\parallel}}
~{\rm{and}}~
\tan\beta=\sqrt{\rho_{tt}\over\displaystyle \rho_{\parallel\parallel}} 
\end{equation}

For Fe and Cu respectively, the values of  $\tan\alpha\ (\tan\beta)$  are found to be 0.562 (0.456) and 0.463 (0.396). This increase is expected, as directionality decreases in the late transition metals \cite{Eberhart:2001,Eberhart:1996c,Eberhart:1996b, Kioussis:2002}, which corresponds to the filling of anti-bonding orbitals. Similarly the topology of the LS phase is identical to the prototype. While the transition from PM to LS shows no topological change, the charge density at the bond CPs changes rapidly. Both the PM and LS phases of fcc Fe have bonds with directionality greater than that of bcc Fe and prototypical fcc topologies. After a transformation to the HS phase, a different picture emerges.

The topology of the spin-majority and total charge density in the HS phase is identical to the PM and LS phases.  Bond directionality is, however, reduced, in accordance with the band theory of ferromagnetism.  Both the total and spin-majority have bond paths with directionality nearly equal to that of bcc Fe, with $\tan\alpha\ (\tan\beta)$ = 0.522 (0.337) and $\tan\alpha^\uparrow\ (\tan\beta^\uparrow)$ = 0.466 (0.402), respectively.  This strong similarity is a result of the filling of both $T_{2g}$ and $E_g$, orbitals by the spin-majority, as in the bcc metal. 

\begin{figure}
\includegraphics{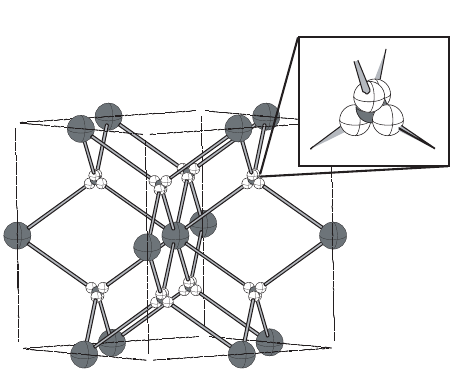} 
\caption{\label{fig:fig3} Topology of the minority spin-density in HS fcc Fe. Only bond points (white) and maxima (gray) are shown for clarity. The appearance of non-nuclear maxima allows the Fe atoms to adopt a bcc-like topology. These replace the cage points found in the tetrahedral holes of the prototype, and the bond points replace the ring points of the prototype. The inset shows the tetrahedron of bond points around a non-nuclear maximum. }
\end{figure}

Interestingly the population of the d-orbitals reducing as $E_g$ in the spin-majority band results in a change in the spin-minority topology. Here, non-nuclear maxima form in the tetrahedral holes. In Fig.~\ref{fig:fig3} these can be seen acting to form a pseudo-bcc topology by allowing each Fe atom to form 8 $\sigma$-bonds. This is due to the fact that  d-orbitals reducing as $T_{2g}$ vastly outnumber those reducing as $E_g$  at moments in excess of 2.5 $\mu_B$. As such, 12 bonds cannot be formed, as that would require both the $T_{2g}$ and $E_g$ representations. In order to stabilize the structure, 8 directional bonds, $\tan\alpha$ = 1.124, are created.

This change in topology is responsible for the magnetovolume instability.  Only two stable topologies exist in fcc Fe:  one corresponds to that observed in the PM-LS phase, while the other is that of the spin-minority band of the HS phase.  The PM/LS (prototypical fcc) topology can be formed so long as $\mu \leq 1.3  \mu_B$. Moments in this regime still allow the spin-minority to use the $E_g$ representation for directional bond path formation. In order for this topology to transform to the HS  topology, a CP annihilation--a topological catastrophe--must occur within the spin-minority topology.  The spin-majority topology, on the other hand, remains unchanged through the phase transition.  

The transformation of the spin-minority topology requires the 12 bond points about each atom in the LS phase to become ring points.  In conjunction, the tetrahedral cage points must transform to non-nuclear maxima.  At the same time, ring points become the bond points shown in Fig.~\ref{fig:fig3}.  During the catastrophe, the curvature at the bond points will become flat in one direction, e.g. a zero eigenvalue of the Hessian (and a vanishing of the Gaussian curvature). Thus, when the population in the $E_g$ representation drops ($1.3 \mu_B < \mu < 2.5 \mu_B$), bond paths in the spin-minority band are weakened and the resultant topology is unstable.  

By way of example, the directionality of the bond points in the minority spin-density, measured from the tetrahedral hole, drops from $\tan\beta^\downarrow$ = 0.380 to 0.240, as the moment is changed from 1.3 $\mu_B$ to 2.1 $\mu_B$ at $a=3.563\AA$. If, however, the moment is allowed to increase to 2.5 $\mu_B$, the number of bond paths is reduced to 8, but their directionality is increased to 1.20. Thus, any moment in the range of $1.3 \mu_B < \mu < 2.5 \mu_B$ will not be observed.  Instead, the topology will spontaneously relax, resulting in a discontinuous change in moment. In a one-electron picture, it is the shift of electrons from the spin-minority to the spin-majority $E_g$ bands that brings about the phase transformation.

In summary, we have applied Bader's AIM theory to the spin-density in bcc and fcc iron. It has been shown that the spin-majority and spin-minority topologies are different in both bcc Fe and the HS phase of fcc Fe. In each case, the two topologies accumulate charge in different regions due to spin-spin correlation.  While the spin-minority density in bcc Fe has cage points between second neighbor atoms, the spin-majority density has bond points.  Here it is the spin-minority topology that resembles the total. In HS fcc Fe this picture is reversed. The spin-majority topology is typical of fcc metals with bonds points lying between the 12 nearest neighbor atoms and cage points in both the tetrahedral and octahedral holes.  The spin-minority topology, however, has cage points in only the octahedral holes. Non-nuclear maxima are present in the tetrahedral holes. These serve to form a pseudo-bcc topology with 8 bond paths per atom.  

Using the fact that the topology can only transform through a catastrophe allows us to discern the two magnetic phase changes in fcc Fe. While describing these using $\mu$ as a local order parameter has offered conflicting results computationally, the topological treatment is unambiguous. A topological phase change does not occur in the PM/LS phase change. The LS/HS transition is, however, accompanied by a catastrophe, and the phase change can be conveniently followed by examining the abrupt change in the Bader bonding graph topology.  We are grateful to the Office of Naval Research for financial support of this work.

\bibliography{Refs2}

\end{document}